# Molecular water accumulation on silica measured by land-contrast interferometry with picometer height resolution

**Xuefeng Wang, Ming Zhao and David D. Nolte***

*Department of Physics, Purdue University, West Lafayette, IN 47907, USA*

We observed water film accumulation on silica surfaces in the air to 2 picometer resolution using optical land-contrast (LC) interferometry. The land-contrast approach is non-destructive and allows real-time measurement of thickness variation of small molecular films on a surface. Water accumulation rates are observed to be different on bare or protein monolayer coated silica surfaces. This suggests that water contributes false signals to mass-sensitive biosensors for protein microarray.

Water is the most fundamental and ubiquitous constituent of biological processes. It participates at all levels of biochemistry, from the dynamics of protein folding, to the chemical preparation of surfaces for immobilization of DNA and antibodies on biosensors. Water plays a particularly crucial role on silica surfaces that are the most commonly adopted platform for gene and protein microarray production. At room temperature, even in the air, water spontaneously deposits on silica surfaces as an ultra thin film and forms hydroxyl groups bound to silicon. These hydroxyl groups participate in the process of silanization which modifies the silica surface and allows proteins or genes to be immobilized covalently to the surface. On the other hand, water film as a dielectric layer may contribute false signals to the mass-sensitive biosensors which detect label-free biomaterials including water on microarray. Therefore quantitative measurement of water amount on biosensors surfaces is important to improve surface chemistry of silica and understand how water affects the accuracy of mass-sensitive biosensors. However, such measurement is a difficult metrology problem because the water film is usually extremely thin and covers the entire surfaces. Although several chemical approaches can be used for water detection on silica [1-3], these methods use chemical reactions and are irreversible and cannot be performed in realtime measurement.

In this work, we introduce the land-contrast BioCD (LC BioCD) which uses self-referencing interferometry to directly and non-destructively sense the thin film deposition on surfaces. We observed water film accumulation on modified silica surfaces in air in realtime with thickness sensitivity down to 2 pm. The principle of land-contrast approach is based on two conjugate-quadrature surfaces on a single substrate. The two surfaces have an equal reflectance so that the initial contrast is zero, while reflectance changes caused by molecular attachment have opposite signs. Thus, the contrast varies as the film deposits on the substrate (shown in Fig.1a). This approach is locally self-referencing and performs common-mode rejection of fluctuations of the light source, making it highly sensitive. We use 140 nm $SiO_2$/Si and 77 nm $SiO_2$/Si as the two conjugate surfaces. The reflection coefficients are respectively $\pm 0.38i$ under normal incidence of 633 nm wavelength, and both reflectances are equal to 0.147. The response for thin film attachment at these reflection coefficients is optimized. When a film with $d$ thickness and $n$ refractive index is applied, the contrast changes by [4]

$$\Delta C = \frac{\Delta R}{R} = \text{Im}\left[\frac{(1+r_1)^2}{r_1} - \frac{(1+r_2)^2}{r_2}\right](n^2-1)\frac{\pi d}{\lambda} \qquad (1)$$

where $r_1$ and $r_2$ are the reflection coefficients of two surfaces and $\lambda$ is the wavelength of the probe. Contrast is defined as $C = (R_1 - R_2)/(R_1 + R_2)$ where $R_1$ and $R_2$ are the reflectance of two surfaces with initial value $R$. $\Delta R$ is the absolute reflectance change on both surfaces due to the film deposition. At 633 nm wavelength, 1 nm water film with $n = 1.33$ causes 0.017 contrast change by calculation. So the conversion factor for water is 0.017 per 1 nm. Effective $n$ of ultra-thin water film varies from 1.33. Maxwell-Garnett theory has been applied to modify $n$ and subsequently the conversion factor which was calculated to be in the range from 0.014 to 0.017 with the water film thickness less than 0.3 nm, the diameter of water molecules. In this article, the presented experiment data of water film thickness was already improved by Maxwell-Garnett modification.

To create the conjugate two surfaces on a single chip, we fabricated spot patterns (to mimic microarray geometry) on silicon thermal oxide chips by photolithography and plasma etching (Fig.1 b). The $SiO_2$ on the land (the background of the substrate) was etched to 77 nm with 140 nm in the mesa (the spot region). The size of the chip is 3×4 $mm^2$. Each chip contains 64 spots and is labeled by a unique barcode. As water accumulates on the chip, the reflectance on the mesa increases while the reflectance on the land decreases with equal value. We acquire the reflectance image of the chips, measure the contrast between the mesa and the land, and then calculate the accumulated water film thickness. To study the water accumulation at different scanning mechanism, two scanning mechanisms are adopted in this article. The spinning disc interferometric (SDI) BioCD scanning system [5] uses 633 nm wavelength laser as probe and an APD (Avalanche photo diode) as detector to acquire the reflectance image of the LC chip by spinning the sample. The other system is Molecular imaging interferometry (MI2) [6] which acquires the reflectance image of the chip under microscope equipped with 630 nm photodiode as light source. Both systems acquire the reflectance image under normal incidence and using common-path configuration. One of the fundamental differences between two scanning systems is that the samples move on SDI system while the sample remains static on MI2 system.

In the first experiment, land-contrast chips were prepared to study the water accumulation on silica at four different chemistries: 1) incubated in 90°C water vapor for 2 hours; 2) dehydrated by baking in 150°C oven for 30 minutes; 3) silanized by soaking in 30 mM chlorodimethyl-octadecylsilane in toluene for 8 hours and becomes hydrophobic: 4) silanized and printed with 2.5 nm protein layer (soaked in 50 μg/ml rabbit IgG solution for 30 min) on the surface. Sample 3 and sample 4 were dried by pure nitrogen stream which is one step of standard protocol of protein array production. SDI scanning system was employed to scan the samples. All chips were attached to the edge of a 100 mm diameter disc and spun at 40π/s angular speed in the air. This condition is equivalent to a 6 m/s air flow. Twenty consecutive scans were performed on four chips with 30 min interval with an overall observation time of ten hours (25$^{o}$C room temperature, 40% relative humidity). After the humidity was changed to 80%, another ten hour consecutive scans were performed. And then changed the humidity to 60% and performed the scans. Before the last round of scans, four chips were baked at 90$^{o}$C for 30 minutes to desiccate the surfaces. The humidity was changed to be 40% in the last scans. The contrast changed as a function of time and was converted to water film thickness presented in Fig. 2a. From the curves, the contrast of the chips slowly increases during the exposure in the air, and the humidity affects the increment rates, and the process of desiccation reduces the contrast significantly. These phenomena suggest that the contrast change of the chips was caused by water accumulation.

The curves demonstrate that these surfaces have different abilities to capture water molecules. In the first round of scans, the dehydrated silica gains 340 pm water, while the silanized (hydrophobic) silica gains 290 pm water. Surprisingly, the vapor-treated silica gains 120 pm water film in the air flow. So the 90$^{o}$C vapor may not make the chip adsorb water to the saturation state in 40% humidity in room temperature. Significantly, the protein-on silica gains 660 pm of water after long-term spinning. Water is captured substantially more by protein than by silica. This result shows that water film can be a potential source of background signal for label-free protein biosensors due to the unbalanced water film deposition on the land and on the protein. The false signal is as large as 660-340 = 320 pm water after exposure to the air in 10 hours. This is equivalent to 220 pm protein signal considering that conversion ratio of protein (0.025/nm) is different from that of water (0.017/nm). The equilibrium time of water adsorption is about 3 hours. The sensitivity of detection is estimated as 2 pm because the standard error based on 64 spots is about 2 pm. In the second phase, all surfaces continued to adsorb water when the humidity was increased to 80%, and the adsorption ceased in the third phase when the humidity was 60%. In the forth phase, after desiccation by baking, all surfaces lost water and then regained water in following 10 hours spinning.

In the second experiment, we compared the water accumulation rate on silica surfaces on different scanning mechanisms. Two protein coated silica was prepared and scanned in parallel to eliminate the discrepancy introduced by other condition variations. One sample was scanned by SDI system which measure reflectance image by spinning the sample. The other sample was scanned by MI2 system which statically measures the reflectance image. The scans of both samples were performed simultaneously and in the same room. The two samples were dried by nitrogen stream and 10 hour scans were performed. From fig. 2b, it is evident that both samples adsorb water while the adsorption rates are different. Kinetic Sample (scanned by SDI system) adsorbs water about three times faster than the static sample. We dried both samples by baking them at 90°C to desiccate the surface and perform scans again. Water continues to accumulate on the surfaces.

In the third experiment, we directly measured the false signal of a protein array contributed by water. A protein array consists of 64 Goat IgG spots was created. The protein array was desiccated by dry nitrogen stream. Consecutive scans were performed immediately in following 10 hours. Average height of protein spots was calculated from each scan. The increment of height was observed and presented in Fig. 3. The initial average thickness is 1.646 nm. After exposure to air for 10 hours, the average thickness became 1.694 nm. It gained about 50 pm thickness due to the exposure in air.

In conclusion, we applied the land-contrast method to quantitative and realtime measurement of water accumulation on silica surfaces in the air with the sensitivity of 2 pm. The binding time to equilibrium is about 3 hours. The amount of adsorbed water is different for protein coated surfaces and bare surfaces and therefore causes a 320 pm (210 pm false increment of protein height using conversion factor of protein) false signal to the protein layer with 2.5 nm original thickness. We tested a protein array exposed in air and acquired 50 pm false signal. This test further verified that the error occurs on a practical protein array. This error source has not gained a general awareness and was not calibrated before. Herein we bring this issue to the attentions of the researchers in the field of mass-sensitive biosensors. To improve the accuracy of mass-sensitive biosensors, the humidity should be well controlled to be a constant. In addition, drying the protein array with pure nitrogen array will desiccate the surfaces and cause errors in the following assay. Therefore, drying the protein array by spinning in the air is recommended. Static scan of the array helps to slow down the water accumulation and therefore reduce the false signal of water.

Land-contrast interferometry is not limited to the study of water binding on silica, but can be applied to other molecular reactions on other substrates. Land-contrast will provide a valuable tool for research on surface chemistry and physics.

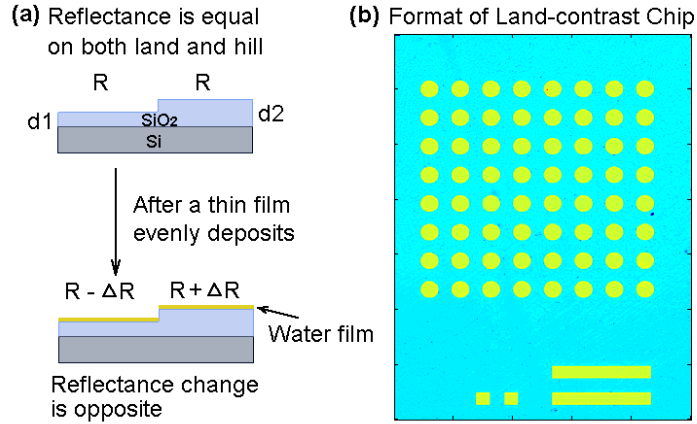

Fig. 1. The principle of the land-contrast interferometry on the BioCD. (a) Reflection coefficients are -0.383i on 77 nm SiO$_2$/Si (land) and 0.383i on 140 nm SiO$_2$/Si (mesa). The reflectance is 0.147 equally. A 1 nm water film causes changes -0.0017 and +0.0017, respectively, on the land and the mesa. (b) Spot patterns were etched on silicon thermal-oxide silicon chips by photolithography. The contrast between the spot and the land is extremely sensitive to water accumulation because the local contrast is self-referenced with common-mode rejection of system fluctuations.

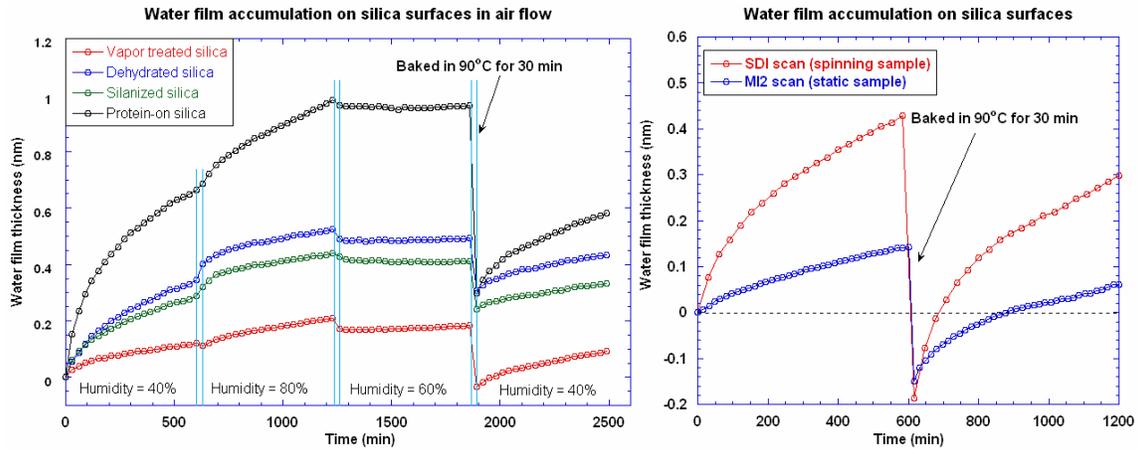

Fig. 2. Water film accumulation on silica surfaces. a). four types of surfaces was consecutively scanned by SDI scanning system on which the samples were spun. The contrast of the samples was converted to the adsorbed water film thickness. The curves demonstrate that these surfaces have different abilities to capture water molecules. Especially the protein coated silica adsorbs more water than the bare silica and this suggests that the water may cause an error to mass-sensitive biosensors for protein arrays. Humidity affects the equilibrium water film thickness. b). water accumulation rate is different on static samples and on samples in motion. Protein coated silica samples were scanned on MI2 and SDI system simultaneously. Water binding rate on sample scanned by SDI is three times higher than on the sample scanned by MI2.

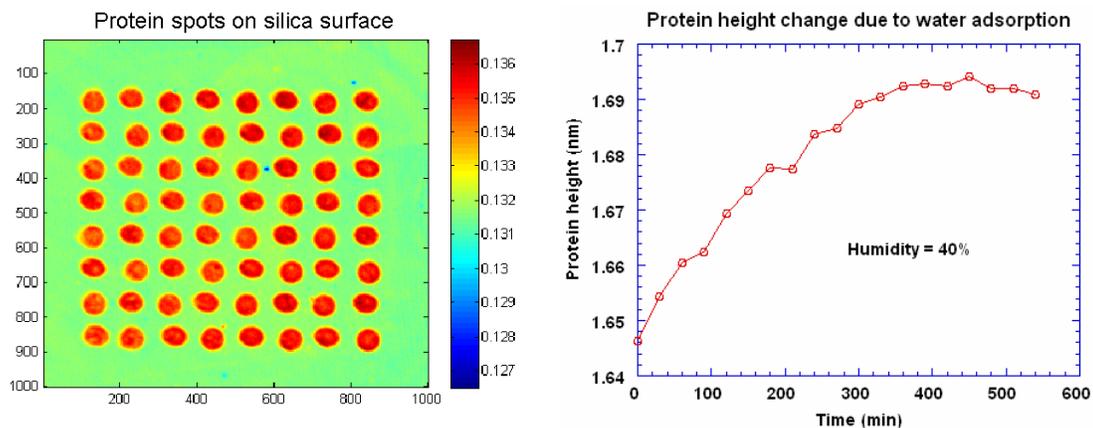

Fig. 3. 64 protein spots (Goat IgG) were printed on 140 nm surface and dried by nitrogen stream. The sample was scanned by SDI system. The measured average protein thickness was continuously increasing in the following 10 hours. About 50 pm error was introduced due to water adsorption imbalance between protein area and the background area.